\documentclass[pra,twocolumn,showpacs,floatfix]{revtex4}
\usepackage{graphicx}
\begin{document}

\title{Solitary waves in mixtures of Bose gases confined in 
annular traps}
\author{J. Smyrnakis$^1$, M. Magiropoulos$^1$, G. M. Kavoulakis$^1$,
and A. D. Jackson$^2$}
\affiliation{$^1$Technological Education Institute of Crete, P.O. 
Box 1939, GR-71004, Heraklion, Greece \\
$^2$The Niels Bohr International Academy, The Niels
Bohr Institute, Blegdamsvej 17, DK-2100, Copenhagen \O, Denmark}
\date{\today}

\begin{abstract}

A two-component Bose-Einstein condensate that is confined in a
one-dimensional ring potential supports solitary-wave solutions,
which we evaluate analytically. The derived solutions are shown
to be unique. The corresponding dispersion relation that generalizes
the case of a single-component system shows interesting features.

\end{abstract}
\pacs{05.30.Jp, 03.75.Lm} \maketitle

{\it Introduction.}
Cold atoms provide an ideal system for the study of nonlinear 
effects including solitary waves. The study of solitary waves 
in these systems has a number of interesting features that 
extend the results of the well-known nonlinear Schr\"odinger 
equation in homogeneous systems, first studied by Zakharov 
and Shabat \cite{ZS} three decades ago.  

Among the many novel aspects associated with the physics of 
solitary waves in trapped gases \cite{Hannover,Phillips,Klaus}
we note (i) the presence of an external trapping potential which 
renders these systems finite, (ii) the fact that they can be 
constructed as quasi-one or quasi-two dimensional systems, and 
(iii) the possibility of bound states of solitary waves in two- 
(or multi-) component gases.  We refer the reader to the article 
by Carretero-Gonzalez {\it et al.}\,\cite{Carr} for a review of the 
extensive work that has been performed on all these problems.

Remarkably, recent experimental advances now permit the realization 
of elongated, quasi-one-dimensional traps.  Solitary waves have been 
observed in elongated traps \cite{Hannover} as well as in more spherical 
traps \cite{Phillips}. More recently, solitary waves have also been 
observed \cite{Klaus} in a two-component Bose-Einstein condensate. 

In the present study we examine the problem of solitary waves 
in a two-component Bose-Einstein condensate 
\cite{Trillo,Chr,Shalaby,Obherg,BA,Frantzeskakis,Berloff,Konotop,Liu} 
on an infinite line. Making an ansatz for the density of the two 
species, we derive exact solutions of the two coupled nonlinear 
equations, which describe the order parameters of the two gases and 
derive the profiles of grey-grey and grey-bright solitary waves, with 
the assumption that the density is constant, either zero or nonzero, 
far from the center of the two waves. We then demonstrate the 
uniqueness of the derived solutions. Finally, we derive the 
dispersion relation associated with these solitary waves, imposing 
periodic boundary conditions, i.e., assuming that the bosons are
confined on a ring. Since the derived solutions are exponentially 
localized, the solutions with periodic boundary conditions can be 
approximated accurately by the present infinite-line calculations 
if the ring is sufficiently large. We compare the resulting 
dispersion relation with that of a single-component Bose gas 
confined in one dimension under periodic boundary conditions 
and interacting via a contact potential, that was first studied 
by Elliott Lieb \cite{Lieb}. 

{\it Model}. 
A Bose-Einstein condensate consisting of two distinguishable
species, $A$ and $B$, is described within the mean-field 
approximation by the two order parameters $\Psi_A(x,t)$ and 
$\Psi_B(x,t)$, which satisfy two Gross-Pitaevskii-like equations,
i.e., two coupled nonlinear Schr\"odinger equations. Assuming 
for simplicity equal masses for the two species, $M_A = M_B = M$, 
and setting $\hbar = 2 M = 1$, these equations have the form
\begin{eqnarray}
 i {\partial \Psi_j}/{\partial t} = 
- {\partial^2 \Psi_j}/{\partial x^2}
+ \gamma (|\Psi_j|^2 + |\Psi_k|^2) \Psi_j,
\label{AB}
\end{eqnarray}
where we have also assumed equal s-wave scattering lengths 
$a_{AA} = a_{BB} = a_{AB} = a$ for all elastic collisions between 
the atoms. The coefficient $\gamma$ that multiplies the nonlinear 
terms is thus equal for all terms and is also $\propto a$. 
In the above equations we have also chosen the normalization 
condition $\int |\Psi_j|^2 \, dx = N_j$, where $N_j$ is the 
number of atoms in each component.

{\it Solitary waves}.
We first determine the solitary wave profiles that result 
from Eqs.\,(\ref{AB}). Writing $\Psi_j(x,t) = \sqrt{n_j(x,t)} 
e^{i \Phi_j(x,t)}$ and separating real and imaginary parts, 
we get two continuity equations for the two species, as well 
as two Euler equations, namely
\begin{eqnarray}
   {\partial n_j}/{\partial t} &+& 2 (n_j \Phi_j')' = 0,
\\
   {\partial {\Phi}_j}/{\partial t}
   = {(\sqrt{n_j})^{''}}/{\sqrt n_j} &-& (\Phi_j')^2
  - \gamma (n_j + n_k),
\end{eqnarray}
where the primes denote spatial derivatives. From the asymptotic 
behavior of Eqs.\,(\ref{AB}), we find that $\Psi_j \propto 
e^{- i \mu t}$, where $\mu = \gamma (n_A^0 + n_B^0)$ is the 
chemical potential and $n_j^0$ is the density of species $j$ 
at $|z| = \pm \infty$.

To derive solitonic solutions, we assume a constant and common 
velocity of propagation $u$ for the two waves, and therefore
we assume that $\Psi_j(x,t) = \sqrt{n_j(z)} e^{i [\phi_j(z) 
- i \mu t]}$, where $z = x - ut$. We note that while 
the density $n_j = |\Psi_j(x,t)|^2$ is indeed a function of 
$z = x - ut$ only, the phase has a more general dependence 
on $x$ and $t$ as indicated. 

{\it Grey-grey solitary waves.}
The assumption of travelling-wave solutions allows us to convert 
the time derivatives into spatial derivatives. The continuity 
equations may then be integrated to give
\begin{eqnarray}
  \phi_j' =  (u/2) (1 - {n_j^0}/{n_j}).
\label{phase} 
\end{eqnarray}
Since the local fluid velocity is $v_j = \hbar \phi_j'/M$, we see 
that $v_j = u (1 - {n_j^0}/{n_j})$. The Euler equations now take the form
\begin{eqnarray}
  \frac {({\sqrt {n_j}})^{''}} {\sqrt {n_j}} = \frac {u^2} 4 
  \left( \frac {(n_j^0)^2} {n_j^2} - 1 \right) 
  + \gamma (n_j - n_j^0 + n_k - n_k^0),
\end{eqnarray}
where we have imposed the boundary conditions $n_j \to n_j^0$ and 
$({\sqrt {n_j}})^{''} \to 0$ for $|z| \to \infty$.

To integrate the above equations, we make the following ansatz, 
which is also consistent with the boundary conditions,
\begin{eqnarray}
  n_B - n_B^0 = \kappa (n_A - n_A^0).
\label{ans}
\end{eqnarray}
This ansatz allows us to perform the integration and 
obtain 
\begin{eqnarray}
   n_A^{'2}/2 &=& [\gamma (1 + \kappa) n_A
- u^2/2] (n_A - n_A^0)^2, 
\label{eqs1}
 \\
   n_B^{'2}/2 &=& [\gamma \frac {\kappa + 1} 
{\kappa} n_B - u^2/2] (n_B - n_B^0)^2.
\label{eqs}
\end{eqnarray}
The consistency of Eqs.\,(\ref{ans}), (\ref{eqs1}), and (\ref{eqs}) 
requires that $n_B = \kappa n_A$, which establishes the value 
of $\kappa$ as $\kappa = n_B^0/n_A^0$ independent of the velocity 
of propagation $u$. Since the sign of $\kappa$ is clearly positive, 
there can be no dark-antidark solitonic solutions that are consistent 
with the ansatz of Eq.\,(\ref{ans}).  In other words, if the $n_j^0$ are 
nonzero, the waves must both be density depressions, i.e., both must 
be grey solitary waves.

The solution of the above equations in the case of grey-grey 
solitary waves has the form
\begin{eqnarray}
 n_j - n_j^0 = - {n_j^0 \cos^2 \theta}/ 
{\cosh^2(z \cos \theta/\xi_D)},
\label{solgr}
\end{eqnarray}
where $\sin^2 \theta = u^2/c^2$ with $c^2 = 2 \gamma 
n_A^0 (1 + \kappa)$ is the (common) speed of sound 
and where the coherence length $\xi_D$ is given by $1/{\xi_D^2} = 
\gamma n_A^0 (1 + \kappa)/2$.

{\it Grey-bright solitary waves.}
The above analysis requires some modifications in the case
where the density of one of the components vanishes at infinity. 
Assuming, for example, that $n_A \to n_A^0 \neq 0$, and $n_B \to 0$, 
the phase of $A$ is still given by Eq.\,(\ref{phase}). However, the 
local velocity of species $B$ is constant and equal to $u$, since 
$\phi_B' = u/2$.

The equations obtained in this case are
\begin{eqnarray}
  \frac {({\sqrt {n_A}})^{''}} {\sqrt {n_A}} = \frac {u^2} 4 
  \left( \frac {(n_A^0)^2} {n_A^2} - 1 \right) 
  + \gamma (n_A - n_A^0 + n_B),
\label{erugb1} \\
  \frac {({\sqrt {n_B}})^{''}} {\sqrt {n_B}} =  
  \gamma (n_A - n_A^0 + n_B) + s,
\label{erugb2}
\end{eqnarray}
where $s$ is the limit of $({\sqrt{n_B}})''/{\sqrt{n_B}}$ 
as $|z| \to \infty$.  The parameter $n_B^0$ in the case of 
grey-grey solutions is now replaced by $s$, which can be 
determined from the number of particles $N_B$ in the bright 
component $B$. The ansatz for the solution assumes the form 
$n_B = \kappa (n_A - n_A^0)$ with $\kappa$ negative. Here, 
$\kappa$ depends on the propagation velocity $u$, in contrast 
to the result obtained above for grey-grey solitons.

The integration of Eqs.\,(\ref{erugb1}) and (\ref{erugb2}) yields 
\begin{eqnarray}
   n_A^{'2}/2 &=& [\gamma (1 + \kappa) n_A
   - u^2/2] (n_A - n_A^0)^2 
\label{grey}
\\
   n_B^{'2}/2 &=&  [\gamma \frac {\kappa + 1} {\kappa} n_B
+ 2 s] n_B^2.
\label{eqs2}
\end{eqnarray}
Consistency of these equations with the ansatz demands 
that 
\begin{eqnarray}
  u^2/2 = \gamma (\kappa + 1) n_A^0 - 2 s,
\label{cond}
\end{eqnarray}
and therefore $\kappa \ge -1$.   Since $\kappa$ must 
also be negative, $-1 \le \kappa \le 0$.

The solution of Eq.\,(\ref{grey}) for the grey component
is the same as in Eq.\,(\ref{solgr}).  The solution 
of Eq.\,(\ref{eqs2}) for the bright component is
\begin{eqnarray}
 n_B = \frac {N_B} {2 \xi_B} \frac 1 {\cosh^2 (z/\xi_B)}
= \frac {-\kappa n_A^0 \cos^2 \theta} {\cosh^2(z \cos \theta/\xi_D)},
\end{eqnarray}
where $\int n_B \, dz = N_B$, in accordance with the 
normalization condition, and $s = 1/\xi_B^2 = \cos^2 \theta/\xi_D^2$.

For completeness we mention that on a ring of finite
radius, our equations also support bright solitary waves in 
both components provided that $\gamma$ is negative. We
will examine this problem in a future study.  

To get some insight into the above results, it is instructive
to write the initial equations taking into account the ansatz,
\begin{eqnarray}
i  \frac {\partial \Psi_A} {\partial t} =
- \frac {\partial^2 \Psi_A} {\partial x^2} +
\gamma [n_A (1+\kappa) + n_B^0 - \kappa n_A^0]
\Psi_A,
\label{anssol1}
\\
i  \frac {\partial \Psi_B} {\partial t} =
- \frac {\partial^2 \Psi_B} {\partial x^2} +  
\gamma [n_B \frac {\kappa + 1} \kappa + n_A^0 - 
\frac {n_B^0} \kappa] \Psi_B.
\label{anssol2}
\end{eqnarray} 
In a sense, the ansatz decouples the two equations, although 
consistency forces additional conditions as shown above. 
The final terms on the right of Eqs.\,(\ref{anssol1})
and (\ref{anssol2}) either vanish for the grey-grey case, 
or are constant for the grey-bright case and are not 
important.  The important terms are those involving 
the combinations $n_A (1 + \kappa)$ and $n_B 
(1 + 1/\kappa)$ or, equivalently, the ``effective" 
couplings $U_A = \gamma (1 + \kappa)$ and $U_B = 
\gamma (1 + 1/\kappa)$. Clearly $U_A = \kappa U_B$. 
In the case of grey-grey solitary waves, $U_A$, 
$U_B$ and $\kappa$ are all positive, which is consistent 
with the fact that the solitary waves are grey in both 
components. On the other hand, in the case of grey-bright 
solitary waves, $U_B$ is negative and $U_A$ is positive 
since $-1 \le \kappa \le 0$. This is consistent with a 
bright wave in the $B$ component and a grey wave in the 
$A$ component. 

{\it Uniqueness of the solutions.} 
It is useful to regard the present problem as that of 
the motion of a particle in a two-dimensional potential,
with $\sqrt{n_j}$ playing the role of spatial coordinates and
$z$ that of time.  For example, in the 
case of two grey solitary waves, 
\begin{eqnarray}
  ({\partial {\sqrt{n_A}}}/{\partial z})^2/2 
+ ({\partial {\sqrt{n_B}}}/{\partial z})^2/2 + V(n_A, n_B) = 0,
\end{eqnarray}
where
\begin{eqnarray}
  V = \frac {u^2} 8 \left[ \frac {(n_A-n_A^0)^2} {n_A} 
+ \frac {(n_B-n_B^0)^2} {n_B} \right]
\nonumber \\  
- \frac \gamma 4 (n_A - n_A^0 + n_B - n_B^0)^2.
\label{pot1}
\end{eqnarray}
In the case of grey-bright solutions the potential becomes,
\begin{eqnarray}
   V = \frac {u^2} 8 \frac {(n_A-n_A^0)^2} {n_A} 
- \frac \gamma 4 (n_A - n_A^0 + n_B)^2 - \frac 1 2 s n_B.
\label{pot2}
\end{eqnarray}
In both expressions for $V$ above, we have imposed the boundary
condition that $V$ vanishes when $n_A$ and $n_B$ have 
their asymptotic values.  

In the grey-grey case, the directional derivative of the potential 
perpendicular to the straight line defined by the ansatz vanishes, 
\begin{eqnarray}
(\sqrt{n_B^0}, -\sqrt{n_A^0}) \cdot \left( \frac {\partial V} 
{\partial \sqrt{n_A}}, \frac {\partial V} {\partial \sqrt{n_B}} 
\right) = 0.
\end{eqnarray}  
Starting at the asymptotic field values at ``time'' $z = -\infty$, 
the system will move along this minimum in $V$ and return to its 
starting point at $z = +\infty$.

The solutions found above for the specific boundary conditions 
are unique. The only physically interesting trajectory must start 
at rest on the $V=0$ contour with the initial values $\sqrt{n_A^0}$ 
and $\sqrt{n_B^0}$ and must end at the same point. Linearization of 
the two coupled nonlinear equations for $|z| \to \infty$ sets the 
asymptotic form of the two solutions. Given this asymptotic form, 
the solutions follow a well-defined path for all $z$ determined by 
the potential of Eqs.\,(\ref{pot1}) and (\ref{pot2}). In other words, 
there is one and only one trajectory which will result from starting 
the system at the point with $(\sqrt{n_A},\sqrt{n_B}) = (\sqrt{n_A^0},
\sqrt{n_B^0})$.

{\it Dispersion relation.}
Having calculated the profiles of the solitary waves, it is
instructive to evaluate the corresponding dispersion relation.
In the case of a single-component Bose gas in one spatial 
dimension, Eliott Lieb \cite{Lieb} found that when bosons 
interact via a contact potential, the excitation spectrum 
consists of two branches. The one corresponds to the usual
Bogoliubov mode, while the other was later shown by Kulish 
{\it et al.} and by Ishikawa and Takayama to correspond to 
solitary waves \cite{Kulish,Ishikawa,KP,JK}. 

\begin{figure}[t]
\includegraphics[width=7.cm,height=4.8cm]{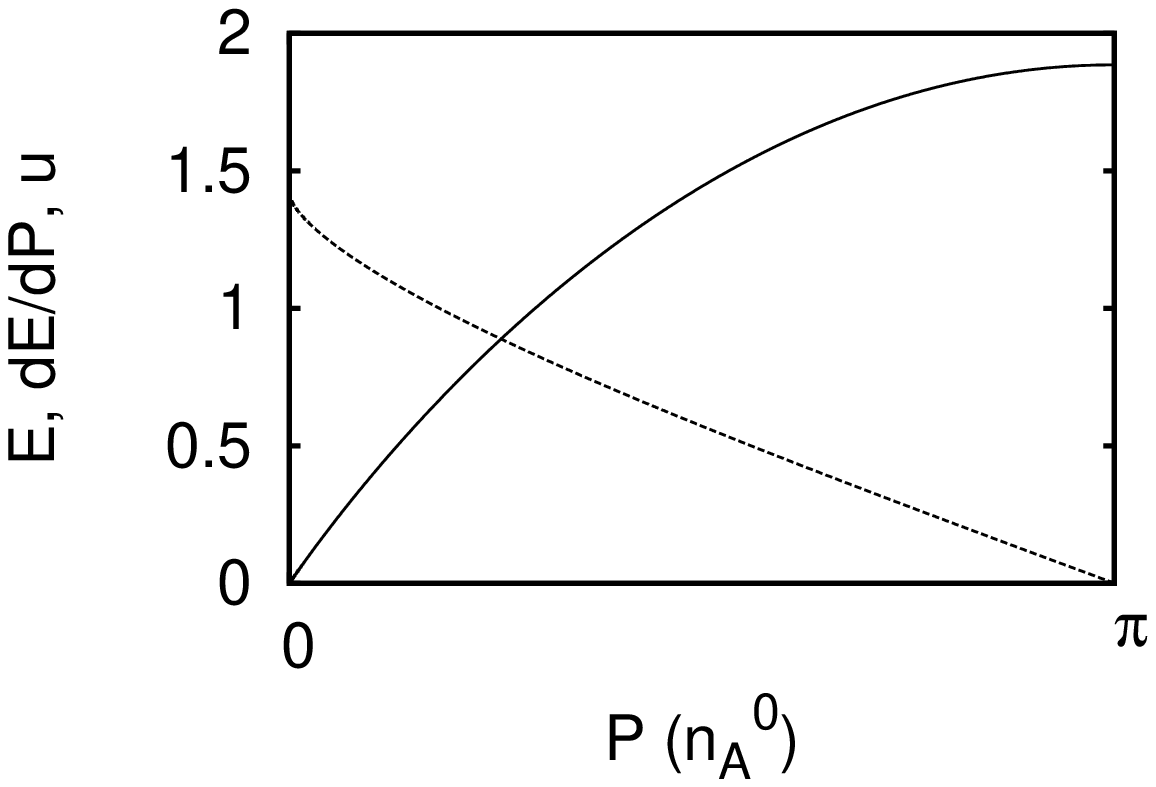}
\includegraphics[width=7.cm,height=4.8cm]{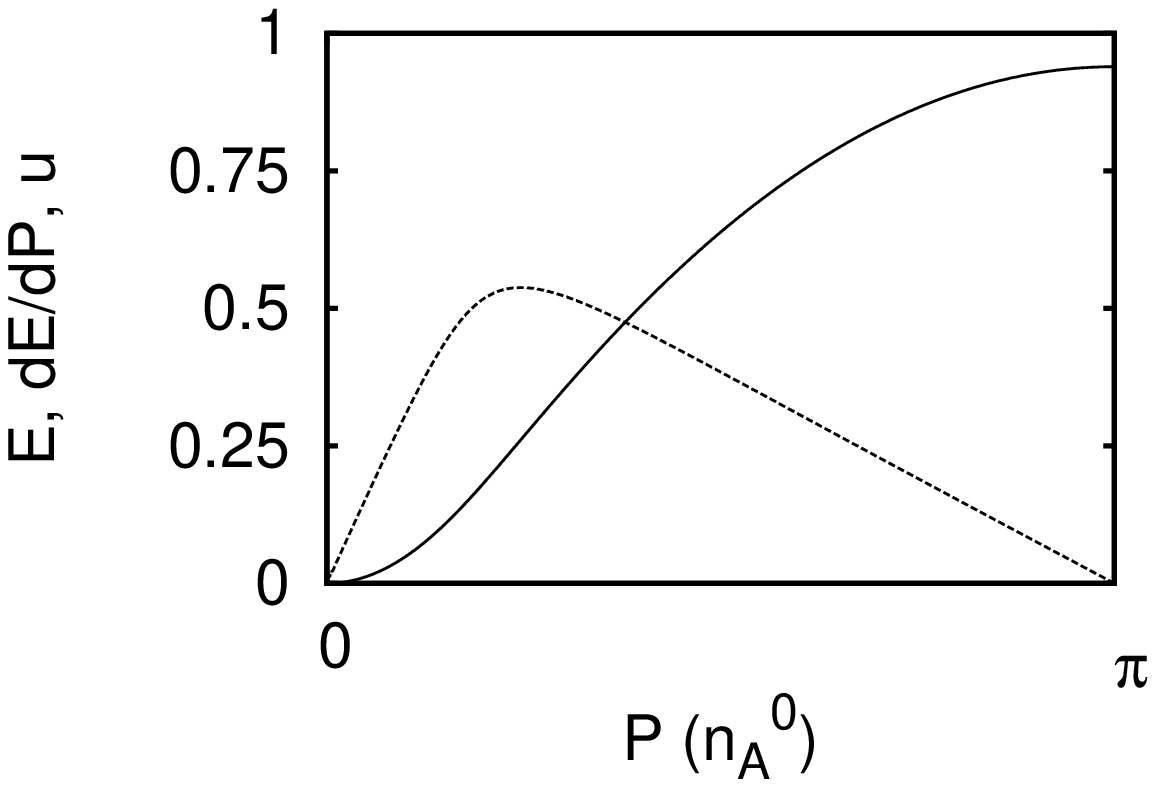}
\caption{The dispersion relation $E=E(P)$ (solid line), 
$dE/dP$ (dashed line), and $u$ (dashed line) as functions 
of $P$ for the case of grey-grey (upper graph) and 
grey-bright (lower graph) solitary waves. The energy 
is measured in units of $E_{GG}$ in the grey-grey case, 
and in $E_{GB}$ in the bright case. The derivative 
$dE/dP$ and $u$ are measured in units $\sqrt{2 n_A^0 
\gamma}$.  The figures show the case $n_A^0 = n_B^0$ 
in the grey-grey case, and $N_B$ is set equal to $n_A^0 
\xi_0$ in the grey-bright case.  The equality $dE/dP = u$ 
is satisfied in both cases.}
\label{FIG1}
\end{figure}

The present calculation generalizes these results to the case
of a two-component system with some similarities and some 
differences. Starting with grey-grey solitary waves, 
the kinetic energy is given by
\begin{eqnarray}
 KE_j = \int [({\partial \sqrt{n_j}}/{\partial z})^2
+ n_i ({\partial \phi_j}/{\partial z})^2] \, dz.
\end{eqnarray}
From Eqs.\,(\ref{phase}) and (\ref{eqs}) the total kinetic 
energy is
\begin{eqnarray}
 KE = KE_A + KE_B = (1 + \kappa)^2 \frac \gamma 2 
 \int (n_A-n_A^0)^2 \, dz.
\end{eqnarray}
The interaction (free) energy less the infinite energy
of the background density of the two components is 
\begin{eqnarray}
 PE - \mu_A N_A - \mu_B N_B 
= \frac \gamma 2 \int (n_A-n_A^0+n_B-n_B^0)^2 \, dz, 
\nonumber \\
\end{eqnarray}
which is equal to $KE$. The total energy is thus
\begin{eqnarray}
 E/E_{GG} = (4/3) (1 + \kappa)^{1/2} \cos^3 \theta,
\label{energy}
\end{eqnarray}
where $E_{GG} = \gamma n_A^0 (1 + \kappa) (n_A^0 \xi_0)$
with $1/\xi_0^2 = n_A^0 \gamma/2$, or $E_{GG} = (1+\kappa) 
[{2 \gamma (n_A^0)^3}]^{1/2}$. Turning to the momentum,
\begin{eqnarray}
 P/n_A^0 =  (1 + \kappa) (\pi - 2 \theta - \sin 2 \theta),
\label{momentum}
\end{eqnarray}
where we have imposed periodic boundary conditions; for a 
detailed analysis of the derivation of Eq.\,(\ref{momentum}),
see Ref.\,\cite{JK}. When Eq.\,(\ref{momentum}) is combined 
with the result of Eq.\,(\ref{energy}), we get the dispersion 
relation $E= E(P)$ plotted in the upper graph of Fig.\,1 for 
the case $n_A^0 = n_B^0$. 

The dispersion relation for grey-bright solitary waves is calculated
in a similar way. The kinetic energy of the two components is
\begin{eqnarray}
 KE_A &=& (\gamma/2) (\kappa+1) \int (n_A - n_A^0)^2 \, dz,
\\
 KE_B &=& \frac 1 2 (n_A^0 \gamma) N_B (1 + \kappa) +
 \frac \gamma 2 \kappa (\kappa+1) \int (n_A - n_A^0)^2 \, dz,
\nonumber \\
\end{eqnarray}
while the potential energy minus the infinite energy due 
to the background density of the $A$ component is
\begin{eqnarray}
 PE - \mu_A N_A = \frac \gamma 2 (1 + \kappa)^2 
\int (n_A - n_A^0)^2 \, dz.
\end{eqnarray}
Collecting terms yields 
\begin{eqnarray}
 E/E_{GB} = \frac 1 2 (1 + \kappa)
\left( 1 - \frac 4 3 \frac {\kappa + 1} \kappa \cos^2 \theta \right),
\label{energygb}
\end{eqnarray}
where $E_{GB} = N_B n_A^0 \gamma$. The momentum of 
the two components is given as \cite{JK} 
\begin{eqnarray}
  P_A / n_A^0 = \pi - 2 \theta - \sin 2 \theta,
\label{nomb0}
\\
 P_B = N_B \sin \theta \sqrt{1 + \kappa} \sqrt{n_A^0 \gamma/2}.
\label{momb}
\end{eqnarray}
The parameter $\kappa$ is now a function of $u$ (and therefore
of $\theta$). Their relationship is established by the
normalization condition,
\begin{eqnarray}
  N_B/(n_A^0 \xi_0) = - 2 \kappa \cos \theta / \sqrt{1 + \kappa}.
\label{normgb}
\end{eqnarray}
From Eqs.\,(\ref{nomb0}), (\ref{momb}) and (\ref{normgb}) we 
find that $P = P_A + P_B$ is given by
\begin{eqnarray}
  P /n_A^0 = \pi - 2 \theta - (1+\kappa) \sin 2 \theta.
\label{momgb}
\end{eqnarray}
Combining Eqs.\,(\ref{energygb}), (\ref{normgb}) and
(\ref{momgb}), we obtain the dispersion relation for a grey-bright
solitary wave, which is plotted in the lower graph of Fig.\,1
for the case $n_A^0 \xi_0 = N_B$. Remarkably, the equation $u = 
\partial E /\partial P$ is satisfied in both graphs. This justifies 
the picture of a solitary wave -- or rather a bound state of two 
solitary waves in the specific case -- as a particle.

While the dispersion relation shown in the upper panel of 
Fig.\,1 is qualitatively the same as in the case of a 
single-component system, with a constant slope for long-wavelength 
excitation (corresponding to sound waves) and zero slope 
at the maximum value of the momentum (corresponding to
dark solitary waves in both components), the lower panel is 
different for long-wavelength excitations. In this limit,
the dispersion relation has a vanishing slope. The limit 
of a static dark solitary wave is qualitatively the same in the
two cases; small amplitude sound waves show a different
behaviour.

Specifically, for the case of grey-grey solitary waves  
when $P \to 0$, we find $E = c P$, with the speed of sound 
$c = \sqrt {2 \gamma (n_A^0 + n_B^0)}$ identical to the value 
obtained by linearizing the original equations \cite{timm}. 
For the case of grey-bright solitary waves when $P \to 0$, the 
dispersion relation vanishes quadratically, $E \to P^2/N_B$. 
Then, $\partial E/\partial P \to \sqrt{2 n_A^0 \gamma (1 + 
\kappa)}$, which agrees with the value of $u$ obtained from 
Eq.\,(\ref{cond}) for $s \to 0$.

{\it Conclusions.} Solitary waves in a two-component Bose-Einstein 
condensate confined in a ring potential show interesting physics. 
Considering repulsive effective interactions, this system supports
bound states of grey-grey and grey-bright solitary-wave solutions, 
which are unique. The dispersion relation resembles that found for 
single-component solitary waves for large density depressions for 
both grey-grey and grey-bright solitary waves. For small density 
depressions, i.e., for sound waves, there is a significant difference 
between grey-grey and grey-bright waves. For long wavelength 
excitation the velocity of propagation $u$ is maximum in the first 
case, which is equal to the speed of sound, and it decreases 
monotonically down to zero when the grey wave become dark, as 
in the case of a single component. On the contrary, in the case 
of grey-bright waves, $u$ vanishes for small density variations, 
i.e., for long-wavelength excitations, and it vanishes also when 
the grey component becomes dark, having a maximum value in between, 
as seen in Fig.\,1.

Experimental verification of the derived dispersion relation shown 
in Fig.\,1 would confirm the expected spectrum for the first time, 
as this has never been confirmed in any nonlinear system so far.
Such an experiment should be possible with use of the method of Bragg 
spectroscopy \cite{Ketterle,Cornell,Ernst}, which has been developed 
to probe the excitation spectrum of cold atomic systems.

\end{document}